\documentclass[ reprint, amsmath,amssymb, aps, ]{revtex4-2}

\usepackage{graphicx} 
\usepackage{hyperref}
\usepackage{xcolor}

\begin{document}
\preprint{APS/123-QED}

\title{Dielectric waveguide setup tested with a superconducting millimeter-wave Fabry-Pérot interferometer at milli-Kelvin temperatures }

\author{J. Lenschen}
\affiliation{Physikalisches Institut, Karlsruher Institut für Technologie}
\author{R. Labbe}
\affiliation{Physikalisches Institut, Karlsruher Institut für Technologie}
\author{N. Drotleff}
\affiliation{Physikalisches Institut, Karlsruher Institut für Technologie}
\author{M. Fuhrmann}
\affiliation{Physikalisches Institut, Karlsruher Institut für Technologie}
\author{J. Lisenfeld}
\affiliation{Physikalisches Institut, Karlsruher Institut für Technologie}
\author{H. Rotzinger}
\email{rotzinger@kit.edu}
\affiliation{Physikalisches Institut, Karlsruher Institut für Technologie}
\affiliation{Institut für Quantenmaterialien und Technologien, Karlsruher Institut für Technologie}
\author{A.V. Ustinov}
\affiliation{Physikalisches Institut, Karlsruher Institut für Technologie}
\affiliation{Institut für Quantenmaterialien und Technologien, Karlsruher Institut für Technologie}

%\date{November 22, 2024}
%\linenumbers   
\begin{abstract}
We propose and test a cryogenic setup comprising dielectric waveguides for mm-wave frequencies in the range of $75-110$\,GHz and temperatures down to 10\,mK. The targeted applications are quantum technologies at millimeter-wave frequencies, which require measurements at low photon numbers and noise. 
We show that the high density polyethylene waveguides combine a frequency independent low photon loss with a very low heat conductance. Black high density polyethylene shows a higher attenuation, which is useful to block thermal photons in a cryogenic environment. 
The dielectric waveguides are thermally anchored and attenuated at several stages of the cryostat. They are individually protected by additional metallic shields to suppress mutual cross-talk and external interference.
With this setup, multiple superconducting resonances of a Fabry-Pérot cavity were measured at 10\,mK. We find quality factors up to 15 million in the single photon limit for resonances above 100\,GHz. These results show no evident influence of atomic two-level systems in the cavity.

\end{abstract}

\keywords{Dielectric waveguide, cryogenic measurement, quantum circuits }
\maketitle

\maketitle

\section{Introduction}

Millimeter-wave measurements at ultra low temperatures are widely unexplored due to several technical difficulties including, for instance, a challenging signal path thermalization and isolation. The tremendous progress in quantum computing at microwave frequencies seen over the last two decades did not so far approach the mm-wave range. The mm-wave quantum measurements would profit from an order of magnitude larger photon energy, wider bandwidth, enhanced resilience against thermal fluctuations, and may speed up qubit manipulation.

The traditionally used metallic hollow waveguides are very good conductors of mm-wave signals but are not flexible and cause an intolerable heat load on the cryogenic temperature stages. Also, the transmission of thermal radiation from elevated temperature stages may be increased due to excess flux of infrared photons through the waveguide, heating up the low-temperature stages. Here, we propose an alternative approach to 100 GHz range quantum measurements by using dielectric waveguides (DWGs) to transmit mm-wave signals from room to milli-Kelvin temperatures and back. DWGs offer several advantages: An ultra low heat conductance, very good transmission properties, and can be easily thermalized. Furthermore, DWGs exhibit very good mechanical flexibility and offer several options to handle the attenuation and suppression of infrared radiation. 
Although it was long known that a dielectric strip can carry mm-waves~\cite{hondros1910ElektromagnetischeWellenDielektrischen}, practical investigations motivated by advances in information technology were carried out only about one decade ago~\cite{Fukuda2011} and are still an attractive area of developments~\cite{dheer2024}. Very recent usage of DWGs for cryogenic applications employed frequencies up to 300\,GHz~\cite{Zhu2020}. There, a circuit board was used to bridge the temperature stages of a cryostat to a base temperature of 4.2\,K.

In this paper, we describe a cryogenic setup including DWG signal lines with high signal attenuation at the input and low signal attenuation at the output lines. We also report on testing the setup with a superconducting Fabry-Pérot interferometer and observation of high-Q resonances in the W-band (75-110\,GHz).  

\section{Ultra-low temperature millimeter-wave setup}

The developed cryogenic mm-wave setup consists of four main parts: A commercial pulse tube pre-cooled dilution refrigerator (Blufors LD250) with a base temperature of 10\,mK, a mm-wave vacuum-transition, black and transparent DWGs bridging the temperature stages and a low noise amplifier with an isolator located at the 2.7\,K stage.  Pre-characterization measurements are also carried out in a home-made dry refrigerator with a base temperature of 2.4\,K (TransMIT 2-stage pulse-tube type PtQUBE).  
The Fabry-Pérot cavity, discussed in the second part of the paper, is directly connected to the DWGs, see Fig.~\ref{fig:cryo_schematic} for a schematic representation. All measurements presented are carried out at room temperature using a Rhode \& Schwarz ZNA67 vector network analyzer (VNA) with external W-band extenders (ZC110).  

\subsection{Vacuum transition}

\begin{figure}[h!]
	\begin{minipage}[c]{0.95\columnwidth}
        \includegraphics[width=\textwidth]{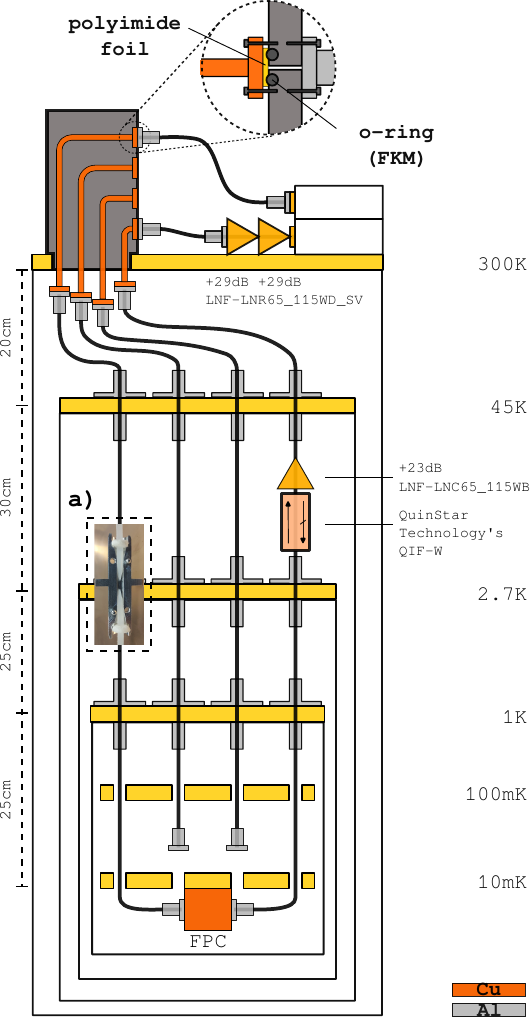}
    \end{minipage}	
    \hfill
    \caption{Schematic of the cryogenic setup with added DWGs. The VNA connect to the four-port vacuum transition (left) which uses a polyimide foil and an o-ring to ensure vacuum tightness. Copper waveguides connect to the DWGs, which are thermally anchored at every stage of the cryostat. The Fabry-Pérot cavity is located at 10\,mK.}
\label{fig:cryo_schematic}
\end{figure}

We use a standard KF40 flange on the top of the fridge to connect a vacuum-tight box equipped with four WR10 vacuum transitions. 
The stainless steel front plate of the box, where the waveguides are connected from the in- and outside using UG-387/UM connectors, is sealed using an o-ring.
On the inside of the box, we use bend copper waveguides. 
We, further, seal each of the spark-eroded rectangular waveguide holes in the metal plate with a $100\,$µm thick polyimide foil and an o-ring, see inset in Fig.~\ref{fig:cryo_schematic} for a schematic. The helium leak rate of the box was measured to be $5\,10^{-8}$\,mbar$\cdot$l/s, confirming the reliability of the polyimide/o-ring seal. A similar approach was previously proposed by~\cite{Koenen2017}.
Transmission measurements reveal an average attenuation of -4\,dB and reflection below -10\,dB over the full spectrum, see Fig.~\ref{fig:room_temp_measurements}. While these values are sufficient for the planned measurements, further optimizations towards increasing the transmission, e.g. by using a thinner polyimide or mica foil, are possible. Also, no special attention was devoted to the suppression of standing waves in the transition plate, see e.g.~\cite{Koenen2017} for a detailed discussion on this matter.

\subsection{Dielectric waveguides}
DWGs, like optical fibers, rely on the principle of total internal reflection which emerges when the carrying medium (the DWG) has a dielectric constant (or refraction index) substantially different from its environment. 
In our approach, we use a rectangular strip of high density polyethylene (HDPE) with a dielectric constant $\epsilon_r = 2.33$ \cite{Macculi2006} embedded in a round polyurethane (PU) foam tube with a diameter of 14\,mm ($\epsilon_r < 1.1$). We have chosen HDPE because of its reported low loss~\cite{weinzierl1998DielectricWaveguidesSubmillimeter}, flexibility at room temperature, low piezoelectricity and low temperature dependence of the dielectric constant. In addition, HDPE material shows good inertness to many chemicals and low moisture sensitivity. 
The DWGs are fabricated from black or transparent HDPE sheets that are cut to a width of 2.5\,mm and then rolled to a thickness of $1.3\,$mm using an in-house made cutting tool, see Fig.~\ref{fig:dwg_cutter}. The electromagnetic properties of the black and transparent DWGs are discussed below.

At mm-wave frequencies, the electromagnetic field of the signal is almost completely confined to the DWG, only a small evanescent field near the DWG's outer boundary is present. According to simulations~\cite{Fukuda2011}, the evanescent electric field amplitude drops of exponentially with about 6\,dB/mm for a material with a comparable $\epsilon_r = 2.6$. DWG materials with higher dielectric constants, like e.g. some glass-reinforced plastics, could be beneficial in terms of a stronger confinement of the electric field. Considering a DWG made from a material with an $\epsilon_r = 6.15$, the surface electric field should decrease by 23\,dB/mm~\cite{Fukuda2011}.

\begin{figure}[h!]
	\begin{minipage}[c]{0.95\columnwidth}
        \includegraphics[width=\textwidth]{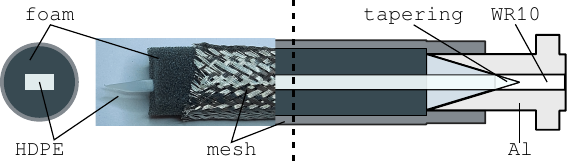}
    \end{minipage}	
	\begin{minipage}[c]{0.95\columnwidth}
        \includegraphics[width=\textwidth]{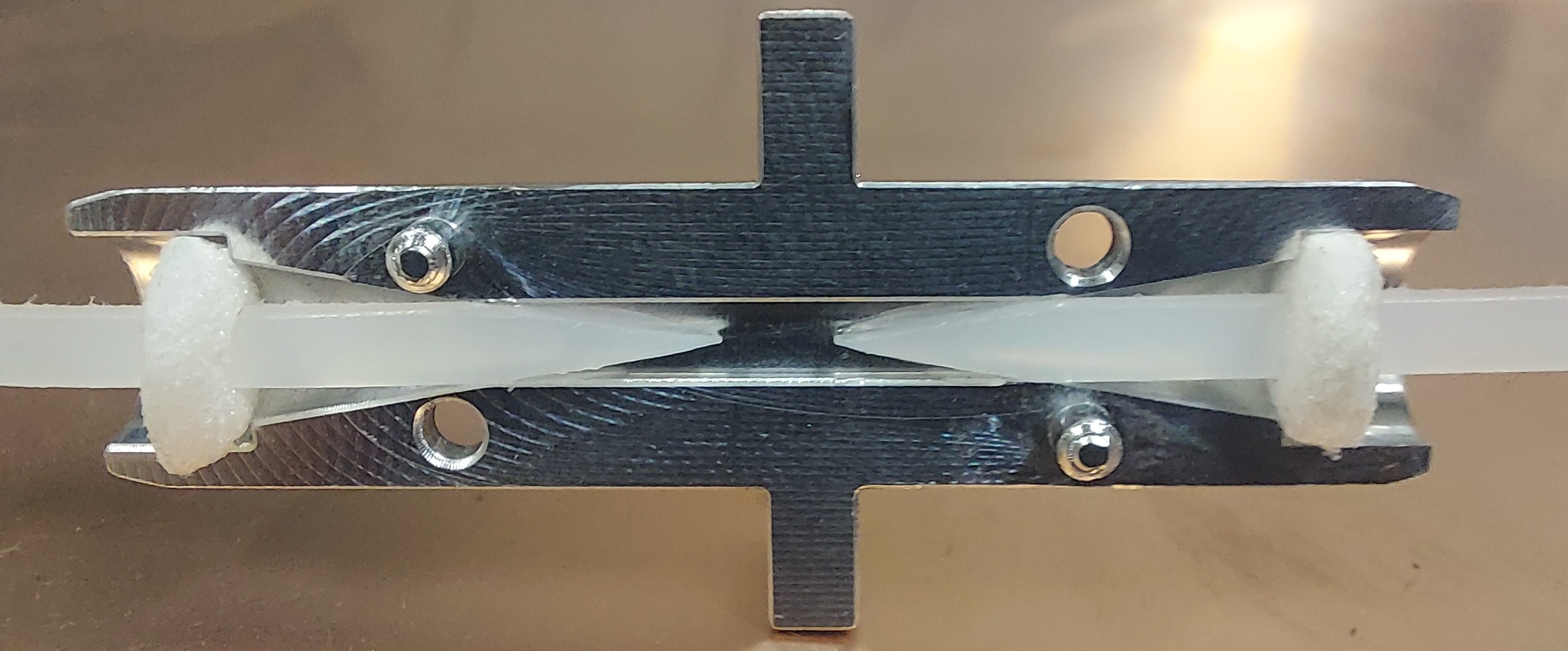}
    \end{minipage}	
\hfill
    \caption{{\bf Top}: Cross sections of the DWG with added aluminum UG-387/UM flange adapter.      
    The image shows a fully assembled DWG with shielding.
    {\bf Bottom}: Picture of a open thermalization transition. The middle part consists of a WR10 waveguide in which the tapered ends of both dielectric waveguides lie. On either side the transition features cone-shaped openings in which two foam disks hold the dielectric waveguides in position while assuring electric decoupling.}
\label{fig:DWG_trans}
\end{figure}

To suppress crosstalk between the lines guiding the signal to and from the low-temperature stage, we place an additional metal shield around the PU foam, see Fig.~\ref{fig:DWG_trans}. This is realized by a metal mesh to retain the flexible character of the DWG.
The shield is especially mandatory at high signal attenuation sections ($ > 30$\,dB), where the higher crosstalk would lead to significant signal distortion without the shielding. A suitable material for the mesh shield is stainless steal, due to its low cryogenic heat conductance, see discussion below. We measure a cross-talk suppression exceeding 60\,dB in the proposed configuration.
Other material solutions for the shield, like stainless steel corrugated tubes or a conductive metallization on the outer side of the PU foam are also possible, but not considered here.

In order to increase signal attenuation and thus thermalization, we make use of the evanescent wave of the DWG by placing a dissipative coating (copper powder filled epoxy) on the dielectric at several positions, see section~\ref{sec:signal_conditioning} for more details.

To connect individual waveguides we developed an adapter with a UG-387/UM flange shown in Fig.~\ref{fig:DWG_trans}. It is constructed out of two cone-shaped aluminum parts that are attached to the tapered ends of the dielectric. By the tapering, a convenient way of impedance matching between the WR10 waveguide and the DWG is provided that can be optimized individually for each transition~\cite{weinzierl1998DielectricWaveguidesSubmillimeter} while the cone-shapes increase coupling to the evanescent portion of the wave, as proposed by \cite{hofmann2003FlexibleLowlossDielectric}. The DWG is held in place by a disk of low loss foam with $\epsilon_r \approx 1.1$ (Millifoam DIV110U).

\subsection{Room temperature properties of dielectric waveguides}

\begin{figure}[h!]
	\begin{minipage}[c]{0.99\columnwidth}
        \includegraphics[width=\textwidth]{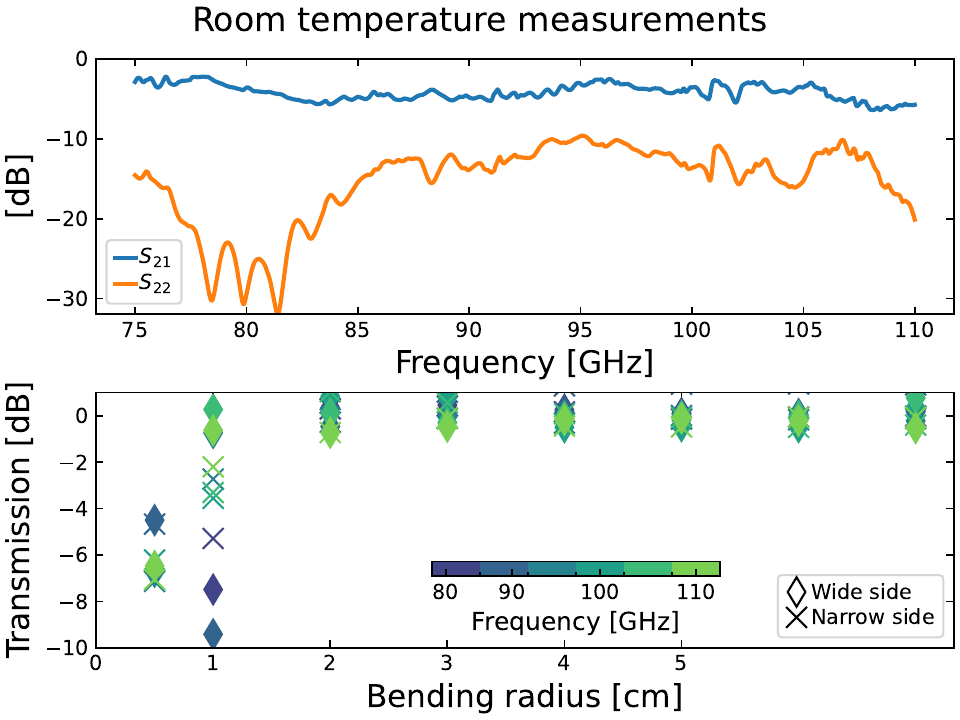}
    \end{minipage}	
    \hfill
    \caption{{\bf Top}: Transmission and reflection measurement of the vacuum transition. {\bf Bottom}: Relative changes in attenuation for a waveguide bent by $180\,^\circ$ with different bending radii. The waveguide under test is made without foam or shielding and consists only of the bare dielectric.}
\label{fig:room_temp_measurements}
\end{figure}
Measurements of the mm-wave transmission through a DWG with varying bending radii show a clear picture, see Fig.~\ref{fig:room_temp_measurements}. Independent of the orientation of the bending, above a radius of $1.5-2\,$cm the transmission remains independent of the bending radius. At smaller radii, the mm-wave loss increases significantly. 

\subsection{Transmission properties of dielectric waveguides at cryogenic temperatures}

\begin{figure}[h!]
	\begin{minipage}[c]{0.99\columnwidth}
        \includegraphics[width=\textwidth]{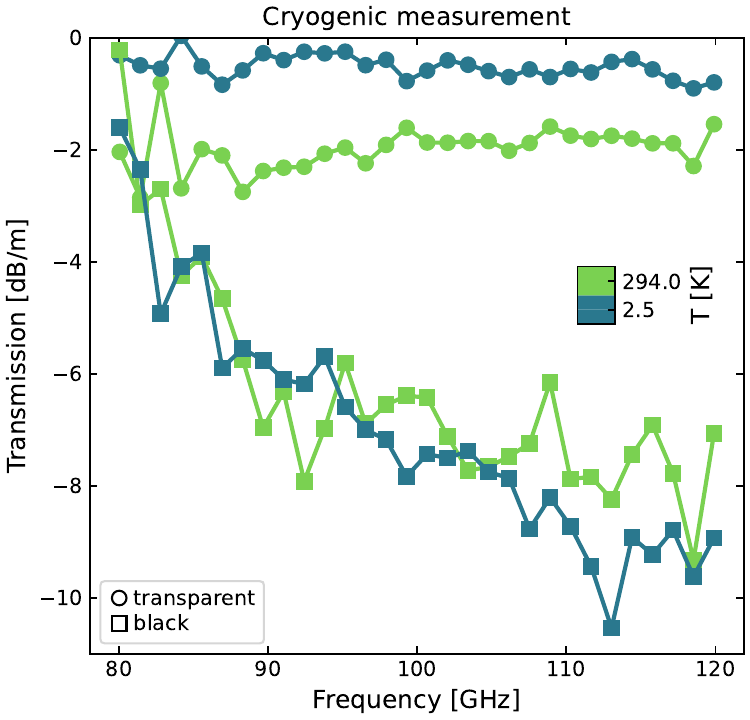}
    \end{minipage}	
\hfill
    \caption{Transmission of a 50\,cm long transparent and a 25\,cm black DWG at room temperature and at $2.5\,$K. The values are normalized to dB/m.}
\label{fig:transmission_v_temp}
\end{figure}

For the cryogenic characterization, we employ 50\,cm (25\,cm) long transparent (black) DWG samples with shielding and UG-387/UM adapters on both sides. Room temperature values are compared with measurements at 2.5\,K. The setup was calibrated in several consecutive cool down steps, with and without DWG samples. The data obtained by this procedure is used to de-embed the cable. Figure~\ref{fig:DWG_trans} shows the resulting transmission data.
The used calibration setting is, however, sensitive to non-stationary influences, e.g. small changes of the DWG-to-adapter impedance matching. We attribute the residual oscillations with an amplitude periodicity of about 10\,GHz to such an origin. Further, noise originating from the pulse tube pre-cooler is observed when measuring the high-Q resonators, see below. This noise is not present when the pre-cooler is switched off, so we suspect that the noise is caused by either vibrational or electrical interference. 

The attenuation of the transparent DWG is roughly frequency independent and decreases with temperature leading to a transmission of about $1.5\,$dB/m at room temperature and 0.6\,dB/m at $2.5\,$K.
Interestingly, the black DWG shows, for frequencies above 95\,GHz, a slightly higher attenuation at low temperature. Below this frequency, the attenuation of the black DWG seems unchanged. The origin of this effect is unknown, but it could hint towards an involvement of carbon as a HDPE filler material, which shows a strong temperature dependence in composite materials, e.g. in the electrical conductance~\cite{pobell1996}. 

The following conclusions can be drawn from above findings. First, both transparent HDPE and black HDPE materials are suitable for low temperature applications. Second, for the transparent DWG, we find a very low cryogenic attenuation, which facilitates measurements of small signals, present for instance in quantum technologies. Third, the stronger damping of the black DWG can be used to attenuate applied mm-wave signals, suppress the background thermal radiation from the high-temperature stages, and thus enhance the noise background at the cold sample under investigation. In addition, the black HDPE filling also reduces the influence of infrared radiation, originating from warmer stages of the cryostat.

\subsection{Dielectric waveguide thermalization}
\label{sec:thermalization}
For thermalization, the DWGs are intersected at the individual stages of the cryostat with metallic transitions using the following approach. The signal is coupled out of the dielectric waveguide into a hollow aluminum rectangular waveguide and vise versa. The coupling is achieved in a similar fashion to that described above for the UG-387/UM adapter. An inside-view of the transition piece, which is made from two identical parts that are screwed together, is shown in Fig.~\ref{fig:DWG_trans}. A small drop of epoxy glue is used to secure the DWGs in position, which also enhances the thermal contact with the aluminum casing. 

To get an estimate for the added heat load on the low temperature stages due to the DWGs, we use Fourier's definition of the heat flux in one dimension
\begin{equation}
Q/A = - \kappa \frac{dT}{dx} \approx -\bar{\kappa} \frac{\delta T}{\delta l} 
\end{equation}
where $A$, $\delta l$ are the cross section and length of the heat conductor, respectively. The values for the average heat transfer coefficient $\bar{\kappa}$ are taken from tabulated data. Due to a lack of specific experimental data, the estimation does not distinguish between different variants of polyethylen (e.g. LDPE or HDPE) or color (transparent or black). We did not consider the detailed temperature gradient but instead use the temperature difference $\delta T$ between the stages. The results are listed in Table \ref{tab:thermaliztion}.
\renewcommand{\arraystretch}{1.3}
\begin{table}[h!]
    \centering
    \begin{tabular}{c|c|c|c|c|c|c}
         cryogenic & $\delta l$ & $\delta T$ & $\bar{\kappa}_{PE}$ & $\bar{\kappa}_{SS}$ & DWG & DWG+S \\ 
         stage & [cm] & [K] & [$\frac{\mu\mathrm W}{\mathrm{cm \ K}}]$ & [$\frac{\mu\mathrm W}{\mathrm{cm \ K}}]$     & Q [$\mu$W]   & Q [$\mu$W]\\
        \hline
        45\,K & 17 & 248 & $5-30\,10^3$ $\dagger$ & $5.2\,10^3$ & 1.4 & 1.8 \\
        2.7\,K & 28 & 42.3 & 100 & $2.1\,10^3$& $4.9\,10^{-4}$ & $1.4\,10^{-2}$   \\
        1\,K & 24 & 1.7 & 20 & $1.5\,10^3$& $4.6\,10^{-6}$ & $4.6\,10^{-4}$ \\
        500\,mK & - & 0.5 &   4 & 700 & $6.5\,10^{-7}$ & $1.5\,10^{-4}$ \\
        100\,mK & 15 & 0.9 & 1.5*& 100*& $2.7\,10^{-7}$  & $2.5\,10^{-5}$ \\
        10\,mK & 10 & 0.09 & 1.0*&40* & $2.9\,10^{-8}$ & $1.6\,10^{-6}$ \\
    \end{tabular}
   
         \caption{Estimated thermal heat load Q by the added DWGs at different stages of the cryostat. The values of the DWG+S include the DWG and the stainless steal mesh, the PU foam ($\rho = 0.02$\,g/cm$^3$) was neglected. The 500\,mK stage is added for reference only, we do not thermalize the DWGs there and at 100\,mK only the outer shield.  
         The tabulated values for stainless steel (A = $4.4\,10^{-6}$\,m$^2$) are taken from \cite{pobell1996,Marquardt2002}, for HDPE (A$_\mathrm{eff}$ = $3.25\,10^{-6}$\,m$^2$) from \cite{Gibson1977,Giles1969,Greig1988}. ($\dagger$) The heat conductance of HDPE depends on the extrusion direction, see \cite{Gibson1977} for details. (*) Denotes extrapolated values.}
    \label{tab:thermaliztion}
\end{table}

The cooling power of today's dilution refrigerators is in the range of Watts for the warmer stages and in the micro-Watt range for colder stages. Based on our measured data, from a thermal point of view, thousands of DWG lines could be easily added to the setup without influencing the individual minimal stage temperature significantly. It is interesting to note, however, that the heat conductance of HDPE is comparable to stainless steel above the 45\,K stage. At lower temperatures, its contribution to the overall heat load of a DWG with shielding is negligible, allowing for very efficient thermal isolation. In comparison to a coaxial solution, there is no inner wire, which is advantageous as well. The metallic body of the transition part can be bolted directly to the low temperature stage, which is a significant simplification especially at mK temperatures. 

\subsection{Signal conditioning}
\label{sec:signal_conditioning}
\begin{figure}[h!]
	\begin{minipage}[c]{0.99\columnwidth}
        \includegraphics[width=\textwidth]{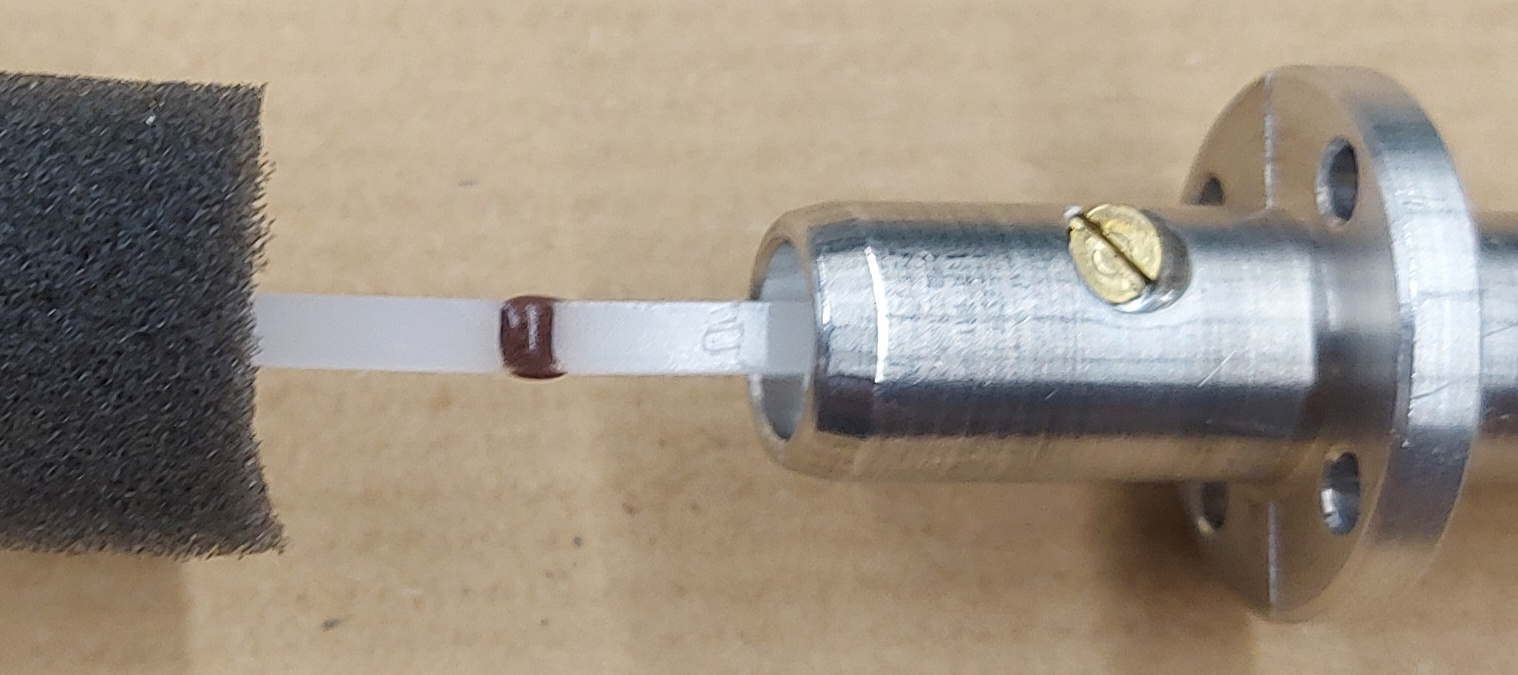}
    \end{minipage}		
\hfill
\caption{Disassembled DWG with $\simeq 1$\,mm copper coating for additional attenuation. The transparent DWG connects to a aluminum stage transition, shown on the right side.}
\label{fig:dwg_copper_coating}
\end{figure}

To thermalize the input signal we used the strategy known from the microwave experiments with superconducting qubits. Using dissipative elements, like the black DWG, the signal as well as the noise is attenuated at cold stages. Thus, the signal to noise ratio remains constant, but the (thermal) Nyquist contribution to the noise drops to the level of the individual temperature stages. Additional attenuation can be obtained by commercial WR10 attenuators, resistive WR10 tubes \cite{Anferov2020} or, as in our case, by adding a dissipative copper powder coating to the surface of the DWG, as shown in Fig.\ref{fig:dwg_copper_coating}. 
To characterize the attenuation, we have measured the transmission (S21) of a transparent DWG with copper coating applied over a length of up to 65\,mm in several steps. These measurements were conducted at room temperature. We observed that the  attenuation is frequency independent and yields a loss of about 1\,dB/mm in the range 85 - 110\,GHz. The reflected signal S11 remained below -20\,dB over the whole frequency range. From the low level of reflection, we conclude that the signal is mostly dissipated by the coupling of the evanescent wave to the copper powder coating. 
The copper coating could also be beneficial in other applications, e.g. to suppress standing waves in the cable that may be caused by imperfect impedance matching at the ends. 

Similar to Refs. \cite{Anferov2020,Suleymanzade2020}, we amplify the signal transmitted through the lowest temperature stage by a factor of about 200 using a commercial wide-band cryogenic high electron mobility (HEMT) amplifier (LNF LNC\_65\_110WB) located at the $2.7\,$K temperature stage. To suppress parasitic resonances as well as back-action from the amplifier on the sample, a WR10 Faraday-type circulator (QuinStar QIF-W) is added at the input of the HEMT. On the room temperature side, we add two amplifiers (LNF-LNR65\_115WD\_SV, +29\,dB each), leading to an overall signal boost of approximately +80\,dB. 
In total, the attenuation from room temperature to the sample on the input line is in the order of 70\,dB, which is sufficient for low photon number operation. 

\section{Fabry-Pérot cavity}

\subsection{Design}
\begin{figure}[h!]
	\begin{minipage}[c]{0.99\columnwidth}
        \includegraphics[width=\textwidth]{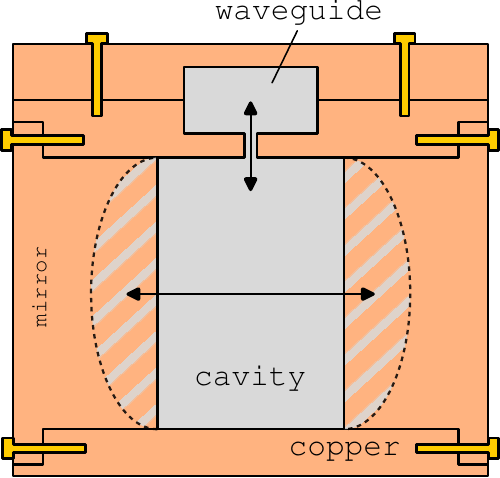}
    \end{minipage}		
\hfill
\caption{Schematic of the Fabry-Pérot cavity (not to scale). Two mirrors (spherical or planar, see text) held in position by a copper support structure. The mm-wave signal of a hollow rectangular waveguide is coupled into the cavity by a hole smaller than the wavelength. Areas in grey depict the hollow volumes while copper parts are colored orange.}
\label{fig:schematic_FPR}
\end{figure}

We follow the general design of Ref. \cite{Kuhr2007} to construct a Fabry-Pérot cavity (FPC) adapted to the frequency range of the setup (W-band), with some important differences discussed below. The FPC consists of two mirrors facing each other with vacuum space in between. We study two configurations, one with two spherical mirrors and another one with a planar and a spherical mirror, see Fig.~\ref{fig:schematic_FPR}. 
The design considerations of a FPC are well known, so we will only briefly review some important points for the two configurations, see e.g.~\cite{Kogelnik1966} for details. The condition for a cavity consisting of two (spherical) mirrors 
\begin{equation}
    0 < \left(1-\frac{d}{R_1}\right) \cdot \left(1-\frac{d}{R_2}\right) < 1
\end{equation}
determines the stability of a standing wave between the mirrors. Here, the distance between the mirrors is given by $d$ while $R_1$ and $R_2$ represent the radii of the curvatures. Following this, one expects resonance frequencies for two curved mirrors with the same curvature when
\begin{equation}
f_{\textit{p,l,q}} = \frac{c}{2d}\left(q+1+ \frac{2p+l+1}{\pi}\arccos\left(1-\frac{d}{R}\right)\right)
\label{f_sperical}
\end{equation}
and for one planar and one curved mirror \cite{Krupka2005} under condition
\begin{equation}
f_{\textit{p,l,q}} = \frac{c}{2d}\left(q+1+ \frac{2p+l+1}{\pi}\arctan\left(\sqrt{\frac{d}{R-d}}\right)\right)
\label{f_planar}
\end{equation}
where $c$ is the speed of light, and $p$, $l$ and $q$ are the mode indices. The terms with the $\arccos$ and $\arctan$ functions are corrections due to the specific geometry of the cavity. The formulas above are related to the unperturbed cavity resonances, i.e. with no external signal coupling or mirror support structures. The coupling scheme, however, can have a strong influence on the individual modes and causes shifts of the observed resonance frequencies. One scheme often employed in the strong coupling regime $Q_c \sim 1000$ makes use of a small hole drilled through one of the mirrors to measure the reflected signal. Since we are interested in coupling quality factors above $Q_c > 100\,000$, we decided for a different approach, as depicted in Fig.~\ref{fig:schematic_FPR}. Here, a WR10 waveguide is passing the cavity body next to the mirrors, with a hole connecting to the cavity. The diameter of the hole is smaller than the corresponding mode wavelength in the cavity. Assuming the hole to be a circular waveguide with a cutoff frequency $f_k \approx 146-160$\,GHz~\cite{meinke1986}, an evanescent wave can enter and leave the cavity with the signal fraction $A/A_0 = e^{\alpha l}$ and the attenuation
\begin{equation}
    \alpha = \frac{2\pi f_k}{c} \sqrt{1-\left(\frac{f}{f_k}\right)^2}.
\end{equation}
With a hole diameter of 1.1-1.2\,mm and a length $l$ of 0.9\,mm this leads to an in-out coupling of about -20\,dB at 100\,GHz. 

Figure~\ref{fig:resonance_fabry-perot} shows a schematic of the FPC. All outer parts as well as the mirror bodies are made in-house from oxygen-free copper. The spherical mirrors have a radius of $25\,$mm and a center-to-center distance of $5.7\,$mm. 

For the fabrication process of the niobium coated mirrors, we have taken the work \cite{Kuhr2007} as a reference. First, the mirrors are machined from an oxygen-free copper block. After careful cleaning, they are annealed at $600\,$°C for $30\,$min in vacuum. Then, the mirrors are polished in several iterations using several diamond based polishing pastes and a stainless steel matrix/pad with the inverse shape. We expect a residual surface roughness in the low µm range. 
Residuals of the polishing paste are removed with acetone and isopropanol. Finally, a 1\,µm thick coating of niobium is deposited in an argon atmosphere by DC-magnetron sputtering under rotation. Without breaking the vacuum, we add a 5\,nm capping of aluminum on top of the niobium coating to prevent oxidation of the niobium.

\subsection{Characterization of the Fabry-Pérot cavity}

One difference of a FPC in comparison to, e.g., a superconducting thin film resonator, is its broad and rich spectrum of modes described by Eqs.~(\ref{f_sperical}) and~(\ref{f_planar}). Our coupling scheme leads to different coupling strengths of individual cavity modes to the waveguide, since it depends on the specific electromagnetic field amplitude in the vicinity of the hole.

\begin{figure}[h!]
	\begin{minipage}[c]{0.99\columnwidth}
        \includegraphics[width=\textwidth]{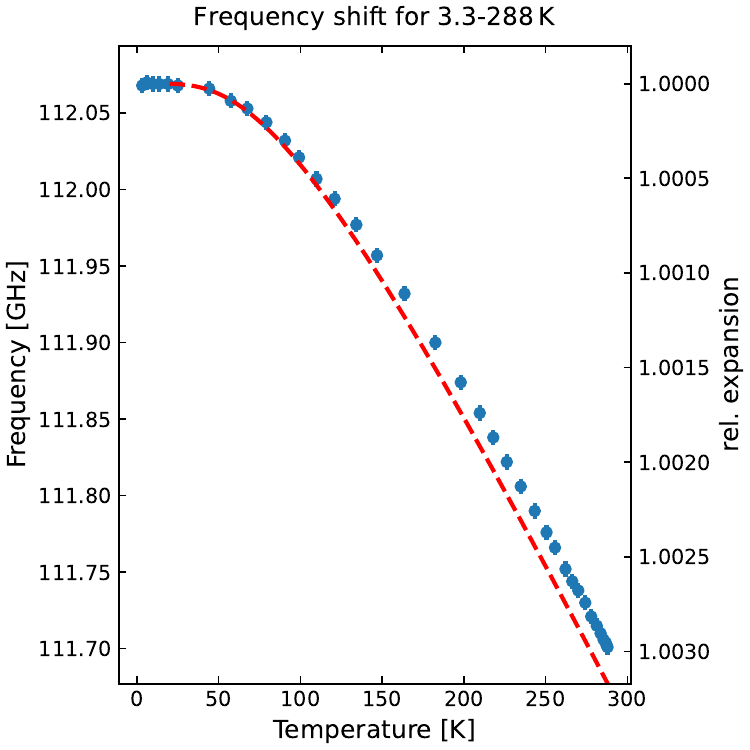}
    \end{minipage}		
\hfill
\caption{Frequency shift of a FPC mode at 112\,GHz between room temperature and 3.3\,K. The red overlay (no fit) shows the relative expansion of copper based on tabulated values~\cite{nist_copper}.}
\label{fig:temp_sweep_rt-2k}
\end{figure} 

Some modes with stronger coupling are already visible at room temperature and can be monitored during cool down and warm up. The shift in center frequency (here around 112\,GHz) with temperature of such a mode is shown in Fig.~\ref{fig:temp_sweep_rt-2k}. We understand this behavior as an isometric thermal expansion of the mirror bodies and the support structure, and therefore a changing distance between the mirror surfaces. This leads to a strong temperature dependence with a shift of $370\,$MHz between 40\,K and room temperature. Figure~\ref{fig:temp_sweep_rt-2k} also shows a comparison with tabulated values for the expansion coefficient of copper. At lower temperatures, the frequency dependence flattens out ($\sim 2\,$MHz shift), as expected.

At temperatures below 8\,K, the superconducting state of the niobium coating (T$_\mathrm c = 9.2$\,K) becomes visible. A multitude of new modes appears with coupling quality factors ranging from $10^5$ to $10^7$. 
\begin{figure}[h!]
   \begin{minipage}[c]{0.99\columnwidth}
        \includegraphics[width=\textwidth]{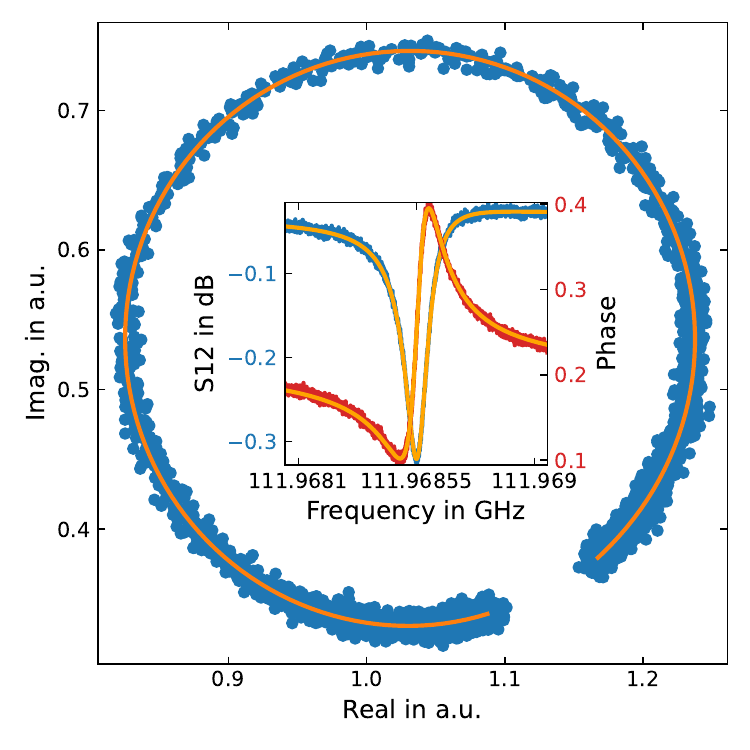}
    \end{minipage}	
\hfill
\caption{The IQ-plane of a resonance measured at 2.4\,K for the FPC is shown, as well as amplitude and phase data in the inset. The internal and coupling quality factor for this particular resonance are $Q_i = 1.2\,10^6$ and $Q_c = 3.7\,10^6$.}
\label{fig:resonance_fabry-perot}
\end{figure} 
A typical resonance curve in the weak coupling regime, measured at 2.5\,K, is shown in Fig.~\ref{fig:resonance_fabry-perot}. The mm-wave response of resonances are analyzed using a circle fit algorithm~\cite{Probst2015} which is implemented in the software package QKIT~\cite{qkit}.

\begin{figure}[h!]
	\begin{minipage}[c]{0.99\columnwidth}
        \includegraphics[width=\textwidth]{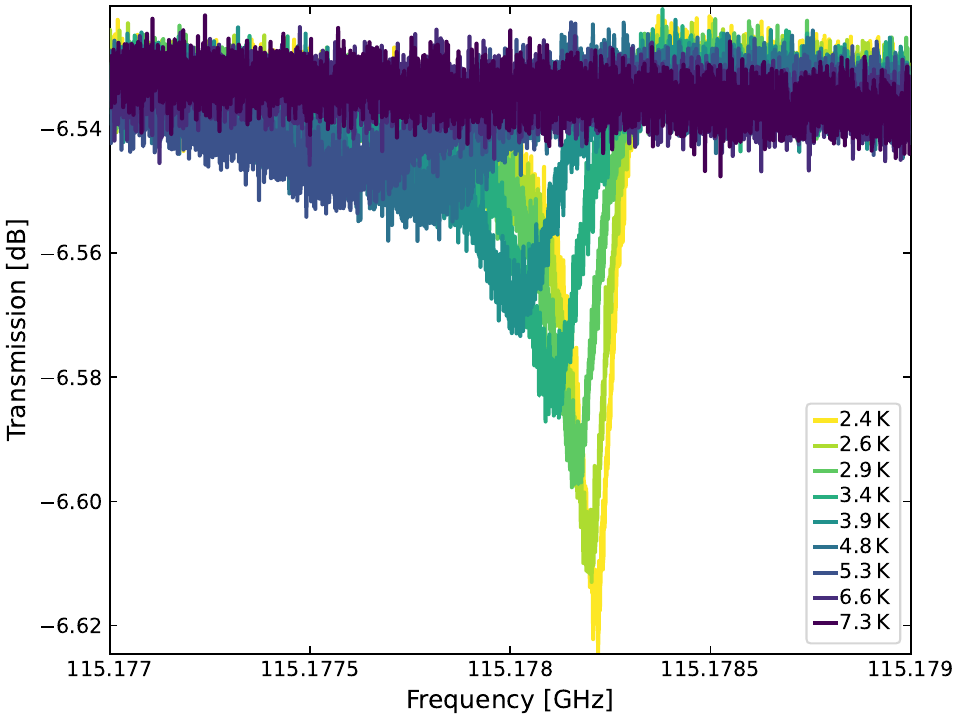}
    \end{minipage}		
\hfill
\caption{Transmission measurement of a resonance at different temperatures.}
\label{fig:resonance_2-8K}
\end{figure}

\begin{figure}[h!]
	\begin{minipage}[c]{0.99\columnwidth}
        \includegraphics[width=\textwidth]{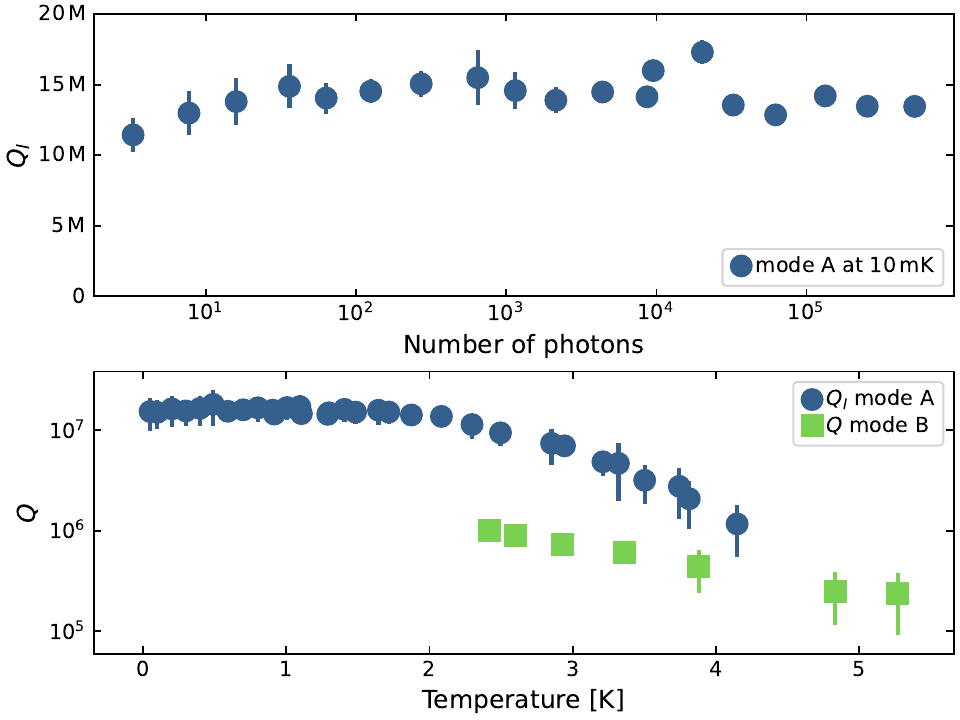}
    \end{minipage}		
\hfill
\caption{{\bf Top:} Power dependence in terms of the average photon number inside the cavity for a mode with an internal quality factor of about $15\,10^6$. This data is taken for a resonance at 107.651\,GHz with coupling quality factors of $46 \pm 3.5\,10^6$. Indicated errors factored by two. {\bf Bottom:} Temperature dependence of the quality factor for two different modes. The blue data points correspond to the mode from the upper plot, while the green data points belong to the mode in Fig.~\ref{fig:resonance_2-8K}. Blue errors factored by 15.}
\label{fig:qi_v_nphotons}
\end{figure}

Figure~\ref{fig:resonance_2-8K} shows the temperature dependence of a resonance curve in the range from 2.5\,K to 6.6\,K, above this temperature the resonance becomes indistinguishable from noise.
With decreasing temperature the resonance frequency shifts to higher values, which we attribute to the appearance of the superconducting energy gap.
 We observe a shift of the resonance dip of about 500\,kHz at 6.6\,K, which allows us to estimate the combined superconducting kinetic inductance fraction and residual thermal expansion of the cavity to be in the order of $10^{-5}$.
 The corresponding change of the quality factor is shown in Fig.~\ref{fig:qi_v_nphotons} (labeled mode B).  

Mounted in the dilution refrigerator setup with a base temperature of 10\,mK, we observe additional modes with higher quality factors. We have investigated two modes ("A" at 107.651\,GHz and "B" at 115.178\,GHz) in detail. The two modes couple to the waveguide sufficiently weak to extract the internal quality factor from the circle fit algorithm. Their corresponding coupling quality factors are $Q_c($A$) = 46 \pm 3.5\,10^6$ and $Q_c($B$) = 14 \pm 0.5\,10^6$.

The power dependence of mode A over a range of 50\,dB, corresponding to a number of photons of about 3 to $10^6$, is shown in the upper part of Fig.~\ref{fig:qi_v_nphotons}. While the internal quality factor decreases slightly from $\sim $15\,M to $\sim $11\,M for photon numbers below 100, the measurement averaging time in this range reaches several hours per data point (photon number), leading to larger systematic errors due to parameter drift. Over the whole range, however, the resonance frequencies lie within an interval of 7\,kHz with a standard deviation of 1.7\,kHz, showing no significant shift.

From these findings, we conclude that there is no apparent evidence for an influence of atomic two-level systems (TLS) on the internal quality factor. For larger photon numbers $\gg 1000$, one expects a significant increase of the internal quality factor in TLS limited resonators \cite{Müller_2019}. 
We attribute the absence of a TLS influence to the large mode volume and thus low electric field density at the surface of the superconducting mirrors of the FPC. 

The temperature dependence of mode A and B is shown in the lower part of Fig.~\ref{fig:qi_v_nphotons}. Measured at an average number of $10^5$ photons, the quality factor of mode A remains constant in the range up to about 2\,K, and decreases to about 1\,M at 4\,K. Mode B shows a similar dependence with an overall lower quality factor. We attribute this behavior to the presence of thermally activated quasi-particles, which finally suppress the modes at 7-8\,K.

\section{Conclusion}

We have demonstrated the feasibility of employing dielectric waveguides for measurements at milli-Kelvin temperatures and low photon numbers. 
The DWGs are simple to manufacture, they are flexible and have low losses as well as a low heat conductance. We have verified options to attenuate applied signals at the cryogenic stages with the use of black DWGs or added copper powder coating. The black DWG also efficiently attenuates infrared radiation. With the presented implementation, thousands of DWGs could be inserted to off-the-shelf dilution refrigerators without exceeding the thermal budget. Additional metallic shields can be used to suppress the electromagnetic cross-talk between signal lines, which is mandatory for measurements at low photon numbers. We verify the performance of the system by characterizing a superconducting Fabry-Pérot cavity at around 100\,GHz, which shows internal quality factors greater than 15\,M and coupling quality factors above 45\,M down to the single photon limit. 

\section{Acknowledgments}

We thank T. Zwick and P. Bushev for fruitful discussions. This work was supported by funding from the European Research Council (ERC) under the European Union's Horizon 2020 research and innovation programme (ERC Advanced Grant project {\em Milli-Q}, grant agreement number 101054327).

\appendix
\counterwithin{figure}{section}
\section{}
\begin{figure}[h!]
	\begin{minipage}[c]{0.99\columnwidth}
        \includegraphics[width=\textwidth]{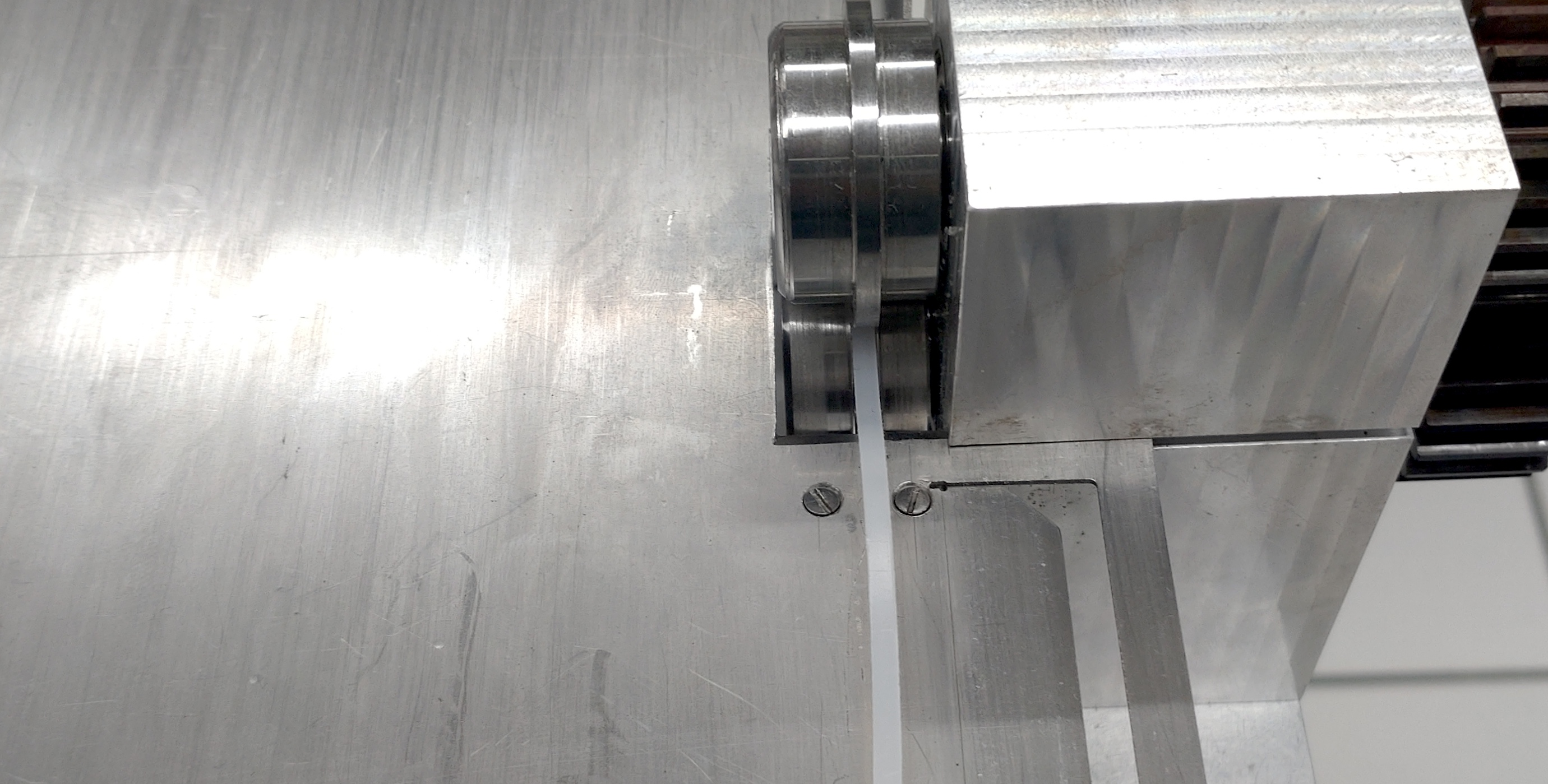}
    \end{minipage}		
\hfill
\caption{Image of the DWG cutting tool. The tool employs two sets of matrices. With the first, we cut the DWG from a 1.5\,mm thick and 2\,m long sheet of HDPE to a width of 2.5\,mm. With the second set of matrices, we roll it to a thickness of 1.3\,mm while maintaining its width. We found that the transmission quality of the signal depends sensitively on the DWG homogeneity as well as the smoothness of the edge, i.e. a sharp cut.}
\label{fig:dwg_cutter}
\end{figure} 

\bibliographystyle{apsrev4-1}
\bibliography{main}

\end{document}